\documentclass{article}
\usepackage{arxiv}

\usepackage[utf8]{inputenc} % allow utf-8 input
\usepackage[T1]{fontenc}    % use 8-bit T1 fonts
\usepackage{hyperref}       % hyperlinks
\usepackage{url}            % simple URL typesetting
\usepackage{booktabs}       % professional-quality tables
\usepackage{amsfonts}       % blackboard math symbols
\usepackage{nicefrac}       % compact symbols for 1/2, etc.
\usepackage{microtype}      % microtypography
\usepackage{lipsum}		% Can be removed after putting your text content
\usepackage{graphicx}
\usepackage[numbers]{natbib}
\usepackage{doi}
\usepackage{soul}
\usepackage{url}
\usepackage{hyperref}
\usepackage{caption}
\usepackage{graphicx}
\usepackage{amsmath}
\usepackage{amsthm}
\usepackage{amssymb}
\usepackage{booktabs}
\usepackage{algorithm}
\usepackage{algorithmic}
\usepackage[switch]{lineno}
\usepackage[capitalize]{cleveref}
\usepackage{multirow}
\usepackage{bbding}

\title{Cap2Sum: Learning to Summarize Videos by Generating Captions}

\author{ {Cairong Zhao\thanks{\quad Corresponding author}} \\
	Department of Computer Science\\
	Tongji University\\
	\texttt{zhaocairong@tongji.edu.cn} \\
	\And
	{Chutian Wang} \\
	Department of Computer Science\\
	Tongji University\\
	\texttt{2030810@tongji.edu.cn} \\
        \And
	{Zifan Song} \\
	Department of Computer Science\\
	Tongji University\\
	\texttt{sugger@tongji.edu.cn} \\
         \And
        {Guosheng Hu} \\
	Oosto, Belfast\\
	\texttt{huguosheng100@gmail.com} \\
	\And
	{Haonan Chen} \\
	Alibaba Group\\
	\texttt{haolan.chn@alibaba-inc.com} \\
        \And
	{Xiaofan Zhai} \\
	Alibaba Group\\
	\texttt{zhaixiaofan.zxf@alibaba-inc.com} \\
}

\renewcommand{\shorttitle}{\textit{arXiv} Template}

\begin{document}
\maketitle
\begin{abstract}
With the rapid growth of video data on the internet, video summarization is becoming a very important AI technology. However, due to the high labelling cost of video summarization, existing studies have to be conducted on small-scale datasets, leading to limited performance and generalization capacity. In this work, 
we introduce the use of dense video captions as a supervision signal to train video summarization models.
%we consider that dense video captions can be the weak labels to train video summarization models. 
%Based on this, 
Motivated by this, we propose Cap2Sum, a model that learns to summarize videos by generating captions, to exploit dense video caption annotations. This weakly-supervised approach allows us to train the models on large-scale dense video caption datasets to achieve better performance and generalization capacity. To further improve the generalization capacity, we introduce a CLIP (a strong vision-language model) Prior mechanism to enhance the learning of important objects that captions may ignore in the videos. In practice, Cap2Sum can perform zero-shot video summarization or be fine-tuned by the ground-truth summary or video caption of the target dataset. To examine the performance of Cap2Sum after weakly-supervised fine-tuning by the video captions, we propose two new datasets, TVSum-Caption and SumMe-Caption, which are derived from 
%. These datasets are based on 
two common video summarization datasets and will be publicly released. We conduct extensive experiments and the results demonstrate that our method achieves significant improvements in performance and generalization capacity compared with previous methods. 
\end{abstract}

% keywords can be removed

\section{Introduction}
\label{sec:intro}
Due to the rapid expansion of video data on the Internet today, video summarization algorithm becomes very useful to users. For example, for those videos recorded by wearable devices that are hours long, it is meaningful to extract the key frames in them. However, the cost of labelling video summarization datasets is extremely high. Specifically, the annotators have to label the importance score for each frame in the videos, and each video requires multiple annotators to reduce subjectivity. Therefore, the existing video summarization datasets (\textit{e.g.}, TVSum \cite{song2015tvsum}, SumMe \cite{gygli2014summe}) are all small-scale datasets, which are insufficient to support the research of video summarization based on deep learning. This fact leads to the limited performance and generalization capacity of existing supervised video summarization 
methods~\cite{park2020sumgraph_sup,gygli2015video_sup,zhang2016video_sup,rochan2018video_sup,zhang2016summary_sup,zhang2018retrospective_sup,wang2024does}. 

To address the insufficiency of large-scale training data, unsupervised and weakly-supervised methods \cite{yuan2019cycle_unsup,mahasseni2017unsupervised,zhou2018deep_unsup,DBLP:conf/eccv/HoCW18_wsup,DBLP:conf/iccv/PandaDWER17_wsup,song2024alchemistcoder,song2024diverse,DBLP:conf/eccv/CaiZDZ18_wsup} have been investigated recently. Generally speaking, previous unsupervised approaches can be divided into (1) adversarial-learning-based methods \cite{apostolidis2020unsupervised_unsup_adv,jung2020global_unsup_adv,jung2019discriminative_unsup_adv,DPL,song2025dpl++} and (2) reinforcement-learning-based methods \cite{zhou2018deep_unsup}. For (1), the video summarization model (\textit{i.e.}, summarizer) is trained to generate a summarization that can reconstruct the original input video. For (2), the summarization is performed by targeting some specific properties (\textit{e.g.}, diversity, length) of an optimal summary. However, for (1), the reconstruction of the original video requires less important information (\textit{e.g.}, backgrounds, environment) contains in the summary, thus is not an ideal goal for training a video summarizer; for (2), 
the optimization targets are very heuristic.
Instead of completely abandoning the usage of any ground-truth labels, weakly-supervised based methods exploit less-expensive weak labels (\textit{e.g.}, video categories, video metadata) to train the summarizers. Unfortunately, due to the insufficiency of the weak labels used, these weakly-supervised methods do not have performance advantages compared to the supervised-based methods. But we still see potential in weakly-supervised methods, as these methods can be trained on large-scale datasets under reasonable supervision.
%the pre-set properties are too rough to train a summarizer. 
% Due to the aforementioned reasons, though the unsupervised models can be trained on large-scale video datasets, existing unsupervised methods do not have competitive performance compared to supervised methods. 

% Recently, weakly-supervised video summarization \cite{DBLP:conf/eccv/HoCW18_wsup,DBLP:conf/iccv/PandaDWER17_wsup,DBLP:conf/eccv/CaiZDZ18_wsup} have achieved promising performance on various datasets. 

We notice that a key factor of the weakly-supervised video summarization is the choice of weak labels. A reasonable label should (a) be easy to labelled, and (b) contain extensive information on video summarization. Based on the above requirements, we ask if we can generate video captions as weak labels to train a strong video summarizer. For (a), video captions are easy to be annotated and therefore many large-scale datasets \cite{msrvtt,youcookii,anet_caption} are currently available. For (b), video captions can also be regarded as the summarization of video contents, but in text form rather than as frame-level scores, making it ideal for training video summarizers. Inspired by this, in this paper, we introduce Cap2Sum, to train video summarizers effectively via adopting video captions as weak labels. Specifically, the proposed model contains two main transformer-based components, a video summarizer, and a video captioner. During training, given a video with video caption annotations, the video summarizer summarizes the video and generates frame-level scores. We use these scores to weigh the video features and feed them to the video captioner to generate dense captions (\textit{i.e.}, text captions with start time and end time). The model is trained and supervised by ground-truth captions. To generate accurate captions, the summarizer is expected to find the frames that are related to the captions, which actually contain the key content of the video, approximating a video summary. In practical, Cap2Sum can generate a summary in a zero-shot setting or be fine-tuned by the labels in the target dataset. The labels employed for fine tuning can either be the ground-truth summary or the video caption. Therefore, to examine the performance of Cap2Sum under the weakly-supervised fine-tuning using video caption, we proposed the TVSum-Caption and SumMe-Caption dataset based on the commonly used video summarization dataset TVSum \cite{song2015tvsum} and SumMe \cite{gygli2014summe}. The proposed datasets will be publicly released for further research. 

In practice, due to the huge domain gap between videos from the video caption datasets and the video summarization datasets (videos from video caption datasets are mostly about the activities of characters and are mostly in third-person view, while videos from video summarization datasets may not contain characters and include first-person videos), the zero-shot performance is not promising for the models pertained on video caption datasets. To address this problem, we propose a CLIP prior mechanism to complement the learning of out-of-distribution objects. This mechanism exploits CLIP \cite{CLIP} to find the frames which contain non-character objects with high information entropy in the videos. We think these frames should be included in the summary and generating auxiliary supervision for training. 
We conduct extensive experiments on TVSum, SumMe, and the proposed datasets. The ablation studies demonstrate the effectiveness of the proposed Cap2Sum architecture and CLIP-Prior mechanism. 
% And our method achieves significant performance improvement compared to other supervised and unsupervised methods (4.5\% and  F1-Score improvement on SumMe using Augmented and Transfer settings respectively compared to the best baseline) .
Our approach achieves significant performance improvement compared to the relevant method CLIP-It \cite{CLIP-It} (4.5\% F1-Score improvement on SumMe using Transfer settings), and reaches state-of-the-art performance among current supervised and unsupervised methods.

Our contribution can be summarized as follows:

\begin{itemize}
    \item We propose Cap2Sum, a model that exploits dense video captions to train a robust and accurate video summarizer.
    \item To further improve the cross-dataset robustness of our model, we propose a CLIP prior mechanism to enhance the learning of the objects that may be ignored by the caption annotations.
    \item We propose two new datasets, TVSum-Caption and SumMe-Caption, for our weakly-supervised fine-tuning. These datasets will be publicly released to support the further research of caption-based video summarization.
    \item We conduct extensive experiments on TVSum, SumMe, and the proposed datasets. The results show that our method achieves significant improvements on performance and generalization capacity compared to previous methods. 
\end{itemize}

\section{Related Work}
\subsection{Video Summarization}
The goal of video summarization is to create a condensed version of the source video that conveys sufficient information and highlights significant events. Current methods can be categorized into three groups: supervised, unsupervised, and weakly-supervised learning approaches.
Supervised learning approaches \cite{zhang2016summary_sup, zhang2016video_sup, zhou2018deep_unsup, zhang2018retrospective_sup, fajtl2018summarizing,  park2020sumgraph_sup, RSGN, CLIP-It, RR-STG, iPTNet,A2Summ} learn from raw videos with corresponding ground truth summary videos annotated by humans. With the help of the benchmark datasets (\textit{e.g.}, SumMe \cite{gygli2014summe} and TVSum \cite{song2015tvsum}), supervised learning approaches have achieved impressive performance. Among them, CLIP-It \cite{CLIP-It} utilizes an off-the-shelf video captioning model and CLIP \cite{CLIP} (a large-scale vision-language model) to generate summary videos conditioned on the text.
iPTNet \cite{iPTNet} employs a joint training approach for both the video summarization and correlated moment localization tasks, which leverages extra moment localization data samples to enhance the performance of video summarization.
A2Summ \cite{A2Summ} builds a unified transformer-based framework for multimodal summarization. 
A significant drawback of supervised video summarization is its reliance on labeled training data. Typically, the datasets used in this field are created by human annotators to access the input video and identify the key frames or shots. Therefore, this highly resource-intensive annotation process leads to the lack of large-scale benchmark datasets for video summarization.

Unsupervised learning approaches \cite{mahasseni2017unsupervised,Cycle-SUM, rochan2018video_sup, song2023SPL,rochan2019video,he2019unsupervised,park2020sumgraph_sup,wu2021era,DSR-RL-GRU,CA-SUM} utilize various hand-crafted heuristics to measure the score of video frames. 
For example, Mahasseni \textit{et al.} \cite{mahasseni2017unsupervised} propose a generative architecture based on variational recurrent auto-encoders and generative adversarial networks to select a subset of key frames. ACGAN \cite{he2019unsupervised} utilizes attentive conditional GANs, with a generator that produces weighted frame features and importance scores, and a discriminator that distinguishes between weighted and raw frame features.
Weakly-supervised learning approaches \cite{DBLP:conf/eccv/HoCW18_wsup,DBLP:conf/iccv/PandaDWER17_wsup,DBLP:conf/eccv/CaiZDZ18_wsup} in video summarization often exploit auxiliary video information such as video categories or metadata. 
Although unsupervised and weakly-supervised learning methods can alleviate the problem of insufficient summary annotations to some extent, it is difficult to design flawless heuristics for the former, while the latter still rely on auxiliary video information, making their performance hard to match that of supervised learning methods.

\begin{figure*}[t]
    \centering
    \includegraphics[width=0.9\textwidth]{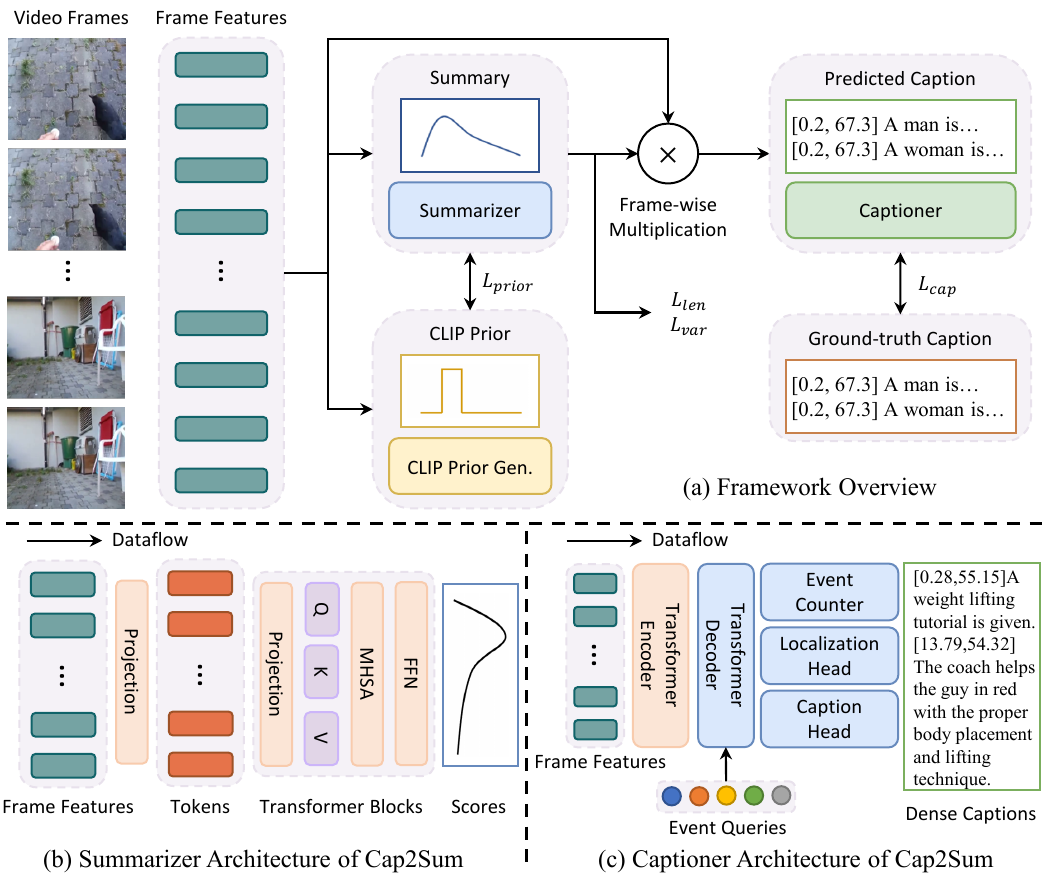}
    \caption{(a) The overview of the proposed Cap2Sum framework. For a input video, the video frames are encoded to features by a pre-trained encoder. The features are fed to the summarizer to generate frame-wise summarization scores. These scores are used to weight the frame features, and are fed to the captioner to generate dense captions. As an auxiliary, a CLIP prior mechanism is proposed to improve the summarization. (b) Architecture of the summarizer in Cap2Sum. (c) Architecture of the captioner in Cap2Sum.}
    
    \label{fig:arch}
\end{figure*}

\subsection{Discussion}
When we train a video summarizer, we actually expect our model to learn the human prior knowledge of what humans consider ``important" in a video. For the supervised learning, this prior knowledge of ``importance" contains in the ground-truth label (\textit{i.e.}, the importance score under human annotation) and is easy to learn. But for unsupervised or weakly-supervised learning, we have to find other ways to let the model learn this human prior. The promising performance improvement achieved by CLIP-It \cite{CLIP-It} has demonstrated that the video captions do relate to the video summarization. 
% Inspired by their success, we can consider the video captions as another type of video summarization in text rather than frame-wise important scores. 
%As evidence, the performance improvement achieved by CLIP-It demonstrates that the video captions do relate to the video summarization. 
However, the improvement in CLIP-It is limited by the accuracy of the generated captions. According to our tests, the caption model BMT \cite{BMT_Iashin_2020} used in CLIP-It performs poorly on the video summarization datasets (\textit{i.e.}, TVSum and SumMe). Besides, the training of CLIP-It is still conducted on the small-scale dataset, leading to the limited zero-shot performance.  %Therefore, we propose to train the video summarization model on the dense video caption datasets and employ the video captions as weak labels instead of extra inputs. 
Generally speaking, compared to CLIP-It, the proposed Cap2Sum has following improvements and novelties: (1) Instead a combining of all the captions in a long video, we propose to use the dense video captions, which are more accurate for the learning of important events in the videos. (2) CLIP-It uses the video captions generated by off-the-shelf models as extra inputs while we adopt the video captions as weak labels, that means the proposed Cap2Sum can be pre-trained on large-scale dense video caption datasets and tested on any videos without any annotation. (3) In our method, the video summarizer can generate captions without video captioner during testing. Thus, the performance of Cap2Sum is not limited by the off-the-shelf pre-trained video captioners. 

\begin{figure*}[t]
    \centering
    \includegraphics[width=0.83\textwidth]{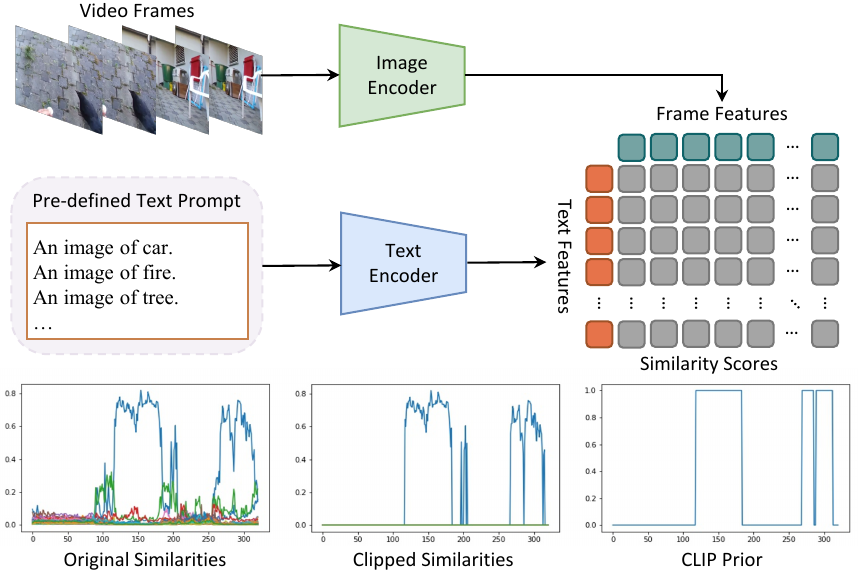}
    \caption{Pipeline of the CLIP Prior Generator. We employ CLIP to calculate the similarity between each frame and pre-defined texts. These similarities are clipped and post-processed to generate the CLIP Prior.}
    \label{fig:clip_prior}
\end{figure*}

\section{Methodology}

As we mentioned in \cref{sec:intro}, the extremely high labelling cost of video summarization motivates us to find other weak labels to train the summarizers. In this section, we propose Cap2Sum, a weakly-supervised video summarization model trained by dense video captions. The major component of Cap2Sum is a video summarizer and a video captioner. Cap2Sum is trained on the dense video caption datasets and expects the model to learn which frames are related to the caption of a video. Through the weakly-supervised training on large-scale datasets, our model achieves strong performance and generalization capacity. 

\subsection{Overview of Cap2Sum}
In this section, we introduce the architecture of Cap2Sum and how model is trained under the supervision of dense video captions.
As shown in \cref{fig:arch} (a), the proposed Cap2Sum is composed of a visual encoder $f_{enc}$, a video summarizer $f_{sum}$, a pre-trained CLIP $f_{clip}$, and a pre-trained video captioner $f_{cap}$. During training, for a given video $\mathbf{V} \in \mathbb{R}^{T\times C \times H \times W}$ with dense caption annotation $c = \{(s_{i},e_{i},t_{i})\}$, the video frames are encoded by an image encoder (we employ the image encoder of CLIP \cite{CLIP} in this paper) to generate frame features $\mathbf{F} = f_{enc}(\mathbf{V})$. A transformer-based video summarizer generates the summary (\textit{i.e.}, frame-wise scores) $\mathbf{S} \in \mathbb{R}^{T}$ based on the visual features. These scores are used to produce weighted features $\mathbf{F}_{w}=\mathbf{S}\cdot\mathbf{F}$, where $\cdot$ denotes element-wise multiplication. Then, the weighted feature is fed to the video captioner to predict captions. As an auxiliary supervision, we propose a CLIP Prior Generator (CLIP Prior Gen.) to generate the CLIP Prior. We introduce this mechanism in detail in \cref{sec:clip_prior}. During testing, for the input videos without annotation, Cap2Sum can perform zero-shot video summarization. To improve the performance, Cap2Sum can also perform few-shot video summarization via the supervised fine-tuning or weakly-supervised fine-tuning on target datasets. Following previous works \cite{CLIP-It,apostolidis2020unsupervised_unsup_adv,jung2020global_unsup_adv}, we use 0/1 knapsack algorithm \cite{Quasi} to select the frames that should be included in the final summary. The details of Cap2Sum are as follows.

% \begin{figure}[t]
%     \centering
%     \includegraphics[width=8cm]{summarizer_arch.png}
%     \caption{Architecture of the summarizer used in Cap2Sum}
%     \label{fig:summarizer}
% \end{figure}

% \begin{figure}[t]
%     \centering
%     \includegraphics[width=8cm]{captioner_arch.png}
%     \caption{Architecture of the captioner used in Cap2Sum}
%     \label{fig:captioner}
% \end{figure}

\noindent \textbf{Video Summarizer.} The key component to perform the summarization is the video summarizer. To generate a reasonable summary, the summarizer has to understand the video contents and combine the context information. Based on the above requirements, we decided to use the video transformer, which has a good capacity to capture global temporal relationships, as the summarizer. As shown in \cref{fig:arch} (b), the frame features are first embedded to the token vectors, and then fed to the transformer blocks. The transformer blocks conduct multi-head self-attention (MHSA) to the tokens. Finally, a MLP head transfer the token sequence to the frame-wise scores.  

\noindent \textbf{Video Captioner.} The captioner employed in Cap2Sum is used to support the training of the summarizer. To obtain the best gradients, we prefer to use a high-performance dense video caption model. In this paper, we apply PDVC \cite{PDVC} as the captioner of Cap2Sum. PDVC is an end-to-end dense video captioner based on deformable transformer. As shown in \cref{fig:arch} (c), PDVC captioner is composed of a transformer encoder-decoder network, an event counter, a localization head and a caption head. The frame features are encoded by the encoder, and then decoded conditioned by the decoder. The event counter predicts the event count in the video, the localization and caption head generates the captions with segments. In practice, the choice of captioner has little impact on the performance of Cap2Sum, and Cap2Sum is able to achieve promising performance on various captioner with acceptable performance.

\subsection{CLIP Prior Mechanism}
\label{sec:clip_prior}
In practice, the videos in the video caption datasets (most are third-person and contain characters) have huge domain gap with the videos in the video summarization datasets (most are first-person and may not contain characters). Besides, the ground-truth captions tend to focus on the activities of the characters and ignore important objects that are not related to the activities. Therefore, when testing on the video summarization datasets under zero-shot setting, the model is hard to correctly label the frames which contain important objects (\textit{i.e.}, fire on the food, bus in the tunnel) as key frames. To address this problem, we propose a CLIP Prior mechanism that exploits the prior knowledge learnt by CLIP \cite{CLIP}. Specifically, CLIP Prior is generated by the CLIP Prior Generator shown in \cref{fig:clip_prior}. We select 100 common object labels mainly based on the categories of CIFAR-100 dataset \cite{cifar100}. We then use CLIP to check whether each object exists in each frame, by calculating the similarity between frames and text. Specifically, we use the prompt ``An image of [object].", where [object] denote the object label, to construct 100 sentences. These pre-defined texts are embedded by the text encoder of CLIP, and then calculate the similarity with the features obtained by the CLIP image encoder (\textit{i.e.}, frame features in \cref{fig:arch}) for each frame via matrix multiplication, formulated as:

\begin{equation}
    \mathbf{M}=softmax\left ( \mathbf{F}_{f}\mathbf{F}_{t}^{\top} \ \right ) 
\end{equation}
where $\mathbf{M}$ denotes the similarity matrix (\textit{i.e.}, the confidence of the specific object appears in the specific frame), $\mathbf{F}_{f}$ is the normalized frame features, and $\mathbf{F}_{t}$ is the normalized text features. We then clip the similarities by threshold $\tau$ to drop uncertain predictions. The clipped similarities are then counted to obtain segments with consecutive values greater than $\tau$, and the length of the segment determines whether the frames corresponding to the segment should be included in the summary. Specifically, for the segments longer than 10 frames and shorter than $0.5T$, we set the corresponding positions in CLIP Prior $\mathbf{P}$ to 1. The other positions in $\mathbf{P}$ are set to 0 and not be used to train the summarizer. The obtained CLIP Prior is used as ground-truth to train the summarizer and encourage it to consider the objects with high information entropy but may not be mentioned in the captions but should be included in the summarization.

\begin{figure*}
    \centering
    \includegraphics[width=0.9\textwidth]{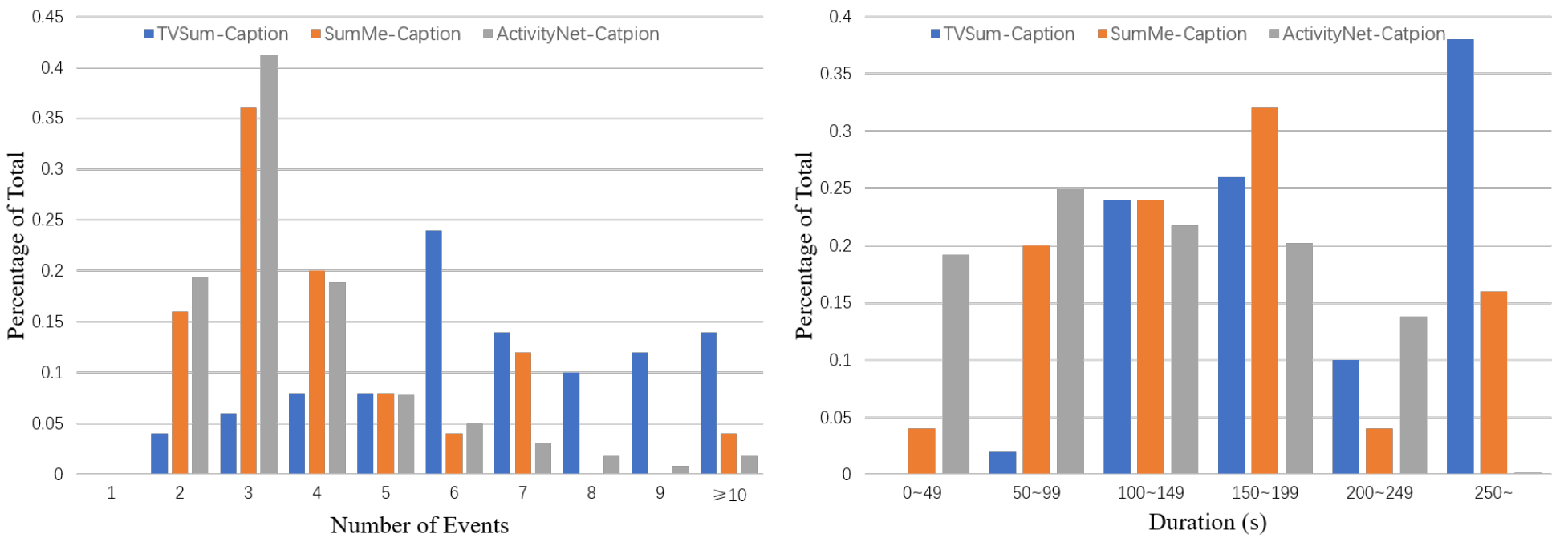}
    \caption{Statistics of the TVSum-Caption dataset and AcitivityNet-Caption dataset.}
    \label{fig:statistics}
\end{figure*}

\subsection{Learning}
The proposed Cap2Sum is trained by a two-stage scheme. First, the model is weakly-supervised pre-trained on the video caption dataset. Then an optional supervised or weakly-supervised fine-tuning is conducted on the video summarization dataset. We next introduce the employed loss functions in detail.

\noindent \textit{A. Weakly-supervised pre-training}

During the pre-training stage, Cap2Sum is trained by 4 loss functions as follows:

\noindent \textbf{Caption Loss.} The major loss function to exploit the caption annotation is the caption loss. Following \cite{PDVC}, the caption loss is composed of $L_{giou}$ (\textit{i.e.}, the generalized IOU \cite{giou} between predicted temporal segments and ground-truth segments), $L_{cls}$ (\textit{i.e.}, focal loss \cite{lin2017focal} between the prediction of prediction scores and the ground-truth), $L_{ec}$ (\textit{i.e.}, the predicted event count and the ground-truth), and $L_{pred}$ (\textit{i.e.}, the cross-entropy between the caption prediction and the ground-truth caption):

\begin{equation}
    \mathcal{L}_{cap}=\beta_{giou}\mathcal{L}_{giou}+\beta_{cls}\mathcal{L}_{cls}+\beta_{ec}\mathcal{L}_{ec}+\beta_{pred}\mathcal{L}_{pred}
\end{equation} \label{equ:loss_cap}
where $\beta_{giou}$, $\beta_{cls}$, $\beta_{ec}$, and $\beta_{pred}$ denote the weights of each loss. 

\noindent \textbf{CLIP Prior Loss.} As mentioned in \cref{sec:clip_prior}, we introduce the CLIP Prior mechanism to enhance the generalization capacity of Cap2Sum. Specifically, we employ a mean-squared error (MSE) loss between the CLIP Prior and masked summarization, formulated as:

\begin{equation}
    \mathcal{L}_{prior}=MSE\left(\mathbf{P} \cdot \mathbf{S},\mathbf{P} \right)
\end{equation}
where the predicted summarization $\mathbf{S}$ is masked by the CLIP Prior to ensure that the positions not mentioned by the CLIP Prior are not considered in the CLIP Prior Loss.

\noindent \textbf{Length Regularization Loss.} To prevent the summarizer generating trivial solutions (\textit{i.e.}, importance scores of all the frames are 1), we employ the length regularization loss to limit the length of the summary, formulated as:

\begin{equation}
    \mathcal{L}_{len}= \left( \frac{\sum_{i=1}^{T}s_{i}}{T} - l \right )^{2}
\end{equation}
where $s_{i}$ indicates $i$-th element in $\mathbf{S}$, and $l$ is the pre-defined target length. In this paper, we set $l$ to 0.3 following previous works. 

\noindent \textbf{Variance Loss.} To enhance the diversity of the generated summarization and avoid over-averaging scores, we introduce the variance loss to maximize the variance of the summarization:
\begin{equation}
    \mathcal{L}_{var}=0.25-\frac{\sum_{i=1}^{T} \left (s_{i}-s_{avg} \right)^{2}}{T}
\end{equation}
where $s_{avg}$ indicates the mean value of the summarization $\mathbf{S}$.

The total loss during pre-training phase is the weighted sum of aforementioned loss functions:

\begin{equation}
    \mathcal{L}_{total}=\beta_{cap}\mathcal{L}_{cap}+\beta_{prior}\mathcal{L}_{prior}+\beta_{len}\mathcal{L}_{len}+\beta_{var}\mathcal{L}_{var}
\end{equation}
where $\beta_{cap}$, $\beta_{prior}$, $\beta_{len}$, and $\beta_{var}$ denote the weights of each loss.

\noindent \textit{B. fine-tuning}

The optional fine-tuning is conducted on the video summarization datasets employed for testing. In this stage, the video summarizer is trained under the supervision of the ground-truth summarization $S_{gt}$ or the captions of the videos. When the ground-truth summarization is employed, the fine-tuning is fully-supervised by the MSE loss between the ground-truth and prediction:

\begin{equation}
    \mathcal{L}_{fine-tune}=MSE\left( \mathbf{S}, \mathbf{S}_{gt} \right )
\end{equation}

However, our method can be fine-tuned by the captions of the videos in the testing dataset. In this scenario, the fine-tuning is weakly-supervised by the caption loss same to \cref{equ:loss_cap}.

\section{The TVSum-Caption and SumMe-Caption datasets}

Since our method can be fine-tuned by the video captions, we need appropriate data (\textit{i.e.}, videos with both caption and summary annotation) to verify the generalization capacity of Cap2Sum under this weakly-supervised fine-tuning scheme. Thus, in this section, we propose the TVSum-Caption and SumMe-Caption datasets. These two datasets are based on the TVSum \cite{song2015tvsum} and SumMe \cite{gygli2014summe} datasets that are commonly used in video summarization research. Specifically, we label the dense video captions for the videos in these two datasets. To generate objective video captions, each video is annotated by 5 annotators, and an experienced supervisor combines the annotations and gives the final annotation. We count the duration and number of events of the videos in the proposed datasets and compare them with the AcitivityNet-Caption (ANet) dataset (i.e the training dataset employed in our experiments), the results are shown in \cref{fig:statistics}. Through the comparison, we notice that compared to the videos in ANet, the videos in TVSum-Caption tend to be longer and contain more events. In addition, although the videos in SumMe-Caption have a similar length and number of events as the videos in ANet, SumMe-Caption contains a large number of first-person videos, which are rarely found in ANet. Therefore, the generalization from ANet to TVSum-Caption and SumMe-caption dataset is challenging. Thus, by conducting experiments on the proposed dataset, we can convincingly verify the generalization capacity of Cap2Sum. The TVSum-Caption and SumMe-Caption dataset will be publicly released to support further research on caption-based video summarization.

\section{Experiments}

\subsection{Settings}
\paragraph{Datasets.} We pre-train Cap2Sum on the ActivityNet-Caption (ANet) dataset \cite{activitynet_caption}. ANet is a large-scale dense video caption dataset that contains over 14000 annotated videos. The fine-tuning and evaluation are conducted on the video summarization datasets TVSum \cite{song2015tvsum} and SumMe \cite{gygli2014summe}. TVSum dataset contains 50 videos, each video has frame-wise importance scores labelled by 20 annotators. SumMe dataset contains 25 videos, and also has frame-wise scores labelled by 15-18 annotators. The proposed caption annotations for TVSum and SumMe datasets are also employed to examine the generalization capacity of Cap2Sum under weakly-supervised fine-tuning. 
%Our experiments include zero-shot settings and few-shot settings. In the zero-shot settings, the model is directly tested without fine-tune. In the few-shot settings, the model is fine-tuned by a few videos in the testing datasets by the ground-truth summary or the text captions.

\paragraph{Data Configuration.} In the previous works \cite{zhang2016video_sup,zhang2016summary_sup,rochan2018video_sup}, the evaluation of video summarization is usually conducted on the TVSum and SumMe datasets in three data settings: Standard, Augment, and Transfer. However, since Cap2Sum requires weakly-supervised pre-training on the large-scale video caption dataset, experiments using Standard setting (\textit{i.e.}, training and testing splits are from the same dataset) are not available. Thus, our experiments are conducted using Augment and Transfer settings. In the Augment setting, the model is pre-trained on the ANet dataset and fine-tuned on the SumMe or TVSum dataset by 80\% of the videos following previous works. In the Transfer setting, the fine-tuning and testing are conducted on the different datasets (\textit{e.g.}, fine-tuning on the SumMe dataset and testing on the TVSum dataset). We also conduct experiments using zero-shot setting (\textit{i.e.}, without fine-tuning) to compare the performance with other unsupervised and weakly-supervised methods. 

\paragraph{Model Configuration.} We employ the pre-trained image encoder (ViT-B/16) of CLIP \cite{CLIP} as the visual encoder in Cap2Sum. The threshold $\tau$ of CLIP Prior mechanism is set to 0.4. $\beta_{giou}$, $\beta_{cls}$, $\beta_{ec}$, and $\beta_{pred}$ are set to 4, 2, 0.5, and 0.5 respectively.  $\beta_{cap}$, $\beta_{prior}$, $\beta_{len}$, and $\beta_{var}$ are set to 2, 10, 0.5, and 0.5 respectively. The model is trained by Adam optimizer \cite{kingma2014adam} with the learning rate set to 5e-5 and batch size set to 1. Following the previous works \cite{zhang2016video_sup,zhang2016summary_sup,rochan2018video_sup,CLIP-It}, we report F1-Score to evaluate the performance of video summarization.

\begin{table}[h]
\centering
\begin{tabular}{@{}l|c|c@{}}
\toprule
Method           & SumMe & TVSum \\ \midrule
unsup.           & 54.5  & 63.2  \\
weak sup.        & 56.9  & 67.4  \\
sup. 5\% videos  & 57.9  & 68.9  \\
sup. 25\% videos & 58.5  & 69.2  \\
sup. 50\% videos & 58.8  & 69.3  \\
sup. 80\% videos & 59.2  & 69.7  \\
\bottomrule
\end{tabular}
\caption{F1-Scores of Cap2Sum fine-tuned by different schemes.}
\label{tab:ablation_fine-tune}
\end{table}

\begin{table}[h]
\centering
\begin{tabular}{@{}cc|c|c@{}}
\toprule
\multicolumn{2}{c|}{Method}                                         & \multirow{2}{*}{SumMe} & \multirow{2}{*}{TVSum} \\
\multicolumn{1}{c}{pre-training} & \multicolumn{1}{c|}{fine-tuning} &                        &                        \\ \midrule
\XSolidBrush                                & \XSolidBrush                                & 53.8                   & 64.0                   \\

\Checkmark                                & \XSolidBrush                                & 55.9                   & 66.8                   \\
\XSolidBrush                                & \Checkmark                                & 54.3                   & 64.7                   \\
\Checkmark                                & \Checkmark                                & 56.9                   & 67.4                   \\ \bottomrule
\end{tabular}
\caption{F1-Scores of removing the proposed CLIP Prior mechanism during pre-training and fine-tuning. $\checkmark$ denotes our CLIP Prior mechanism is employed in the corresponding phase.}
\label{tab:ablation_clip_prior}
\end{table}

\begin{table*}[h]
\centering
\small
\setlength\tabcolsep{8pt}
\begin{tabular}{@{}l|c|cc|cc|c@{}}
\toprule
\multirow{2}{*}{Method}                                     & \multirow{2}{*}{Publication} & \multicolumn{2}{c|}{SumMe} & \multicolumn{2}{c|}{TVSum} & \multirow{2}{*}{Rank} \\
                                                            &                              & Augmented    & Transfer    & Augmented    & Transfer    &                               \\ \midrule
SumTransfer   \cite{zhang2016summary_sup}         & CVPR 2016                                         & 41.3          & 38.5          & -             & -     & 17        \\
dppLSTM   \cite{zhang2016video_sup}               & ECCV 2016                                         & 42.9          & 41.8          & 59.6          & 58.7     & 16     \\
SUM-GAN$_{sup}$   \cite{mahasseni2017unsupervised} & CVPR 2017                                         & 43.6          & -             & 61.2          & -       & 14      \\
SUM-FCN   \cite{rochan2018video_sup}              & ECCV 2018                                         & 51.1          & 44.1          & 59.2          & 58.2     & 11     \\
SUM-DeepLab   \cite{rochan2018video_sup}          & ECCV 2018                                         & 50.2          & 45.0          & 59.1          & 57.4      & 12    \\
DR-DSN   \cite{zhou2018deep_unsup}                & AAAI 2018                                         & 43.9          & 42.6          & 59.8          & 58.9      & 13    \\
$re$-SEQ2SEQ   \cite{zhang2018retrospective_sup}  & ECCV 2018                                         & 44.9          & -             & 63.9          & -          & 9   \\
VASNet   \cite{fajtl2018summarizing}               & ACCV 2018                                         & 51.1          & -             & 62.4          & -         & 8    \\
UnpairedVSN   \cite{rochan2019video}               & CVPR 2019                                         & 48.0            & 41.6          & 56.1          & 55.7     & 15     \\
SumGraph   \cite{park2020sumgraph_sup}            & ECCV 2020                                         & 52.9          & 48.7          & 65.8          & 60.5      & 6    \\
RSGN   \cite{RSGN}            & TPAMI 2021                                         & 45.7          & 44.0          & 61.1          & 60.0       & 10   \\
CLIP-It \cite{CLIP-It}                             & NeurIPS 2021                                      & 56.4          & 51.9          & 69.0          & 65.5   & 3       \\ 
RR-STG \cite{RR-STG}                             & TIP 2022                                      & 54.8          & 45.4          & 63.6          & 59.7    & 7      \\ 
iPTNet \cite{iPTNet}                             & ECCV 2022                                      & 56.9          & 49.2          & 64.2          & 59.8    & 5      \\ 
A2Summ \cite{A2Summ}                             & CVPR 2023                                      & -          & 55.0          & -          & 63.4     & 4     \\ \midrule
Cap2Sum$_{wsup}$                                                    & Proposed                                          & 56.9          & 55.3          & 67.4          & 64.8   & 2       \\
Cap2Sum$_{sup}$                                                     & Proposed                                          & \textbf{59.2} & \textbf{56.4} & \textbf{69.7} & \textbf{66.2} & \textbf{1}  \\ \bottomrule
\end{tabular}
\caption{The comparison of Cap2Sum to supervised approaches using different training settings (\textit{i.e.}, Augmented and Transfer) on SumMe and TVSum, evaluated by F1-Score (\%).}
\label{tab:sup_comp}
\end{table*}

\subsection{Ablation Studies}
\label{sec:ablation_study}
We first conduct ablation studies to demonstrate the effectiveness of the weakly-supervised fine-tuning and the CLIP-Prior mechanism. The ablation studies conducted in this section use the Augment setting (\textit{i.e.}, model pre-trained on ANet is fine-tuned and tested on the same dataset).

\paragraph{The Impact of Fine-tuning Scheme.} 
In this subsection, we conduct experiments to verify the effectiveness of weakly-supervised fine-tuning (\textit{i.e.}, supervised by the dense video caption) on the proposed TVSum-Caption and SumMe-Caption datasets. We also conduct ablation studies on the training split ratio of supervised fine-tuning to verify the few-shot performance of Cap2Sum. We report the F1-Scores of unsupervised setting (unsup.), weakly-supervised setting (weak sup.), and supervised setting using $p$\% videos to fine-tune (sup.  $p$\% videos) in \cref{tab:ablation_fine-tune}. The results show that using video captions to fine-tune Cap2Sum is very effective (gains 2.4\% improvement on SumMe and 4.2\% improvement on TVSum compared to unsupervised setting). This means that the user only needs to provide part of the video captions of the videos that need to be summarized, and Cap2Sum can be effectively generalized to that user's videos. In addition, the comparison of different training split ratios of supervised fine-tuning demonstrates Cap2Sum can achieve promising performance using a few videos to fine-tune. Thus, it is proved that Cap2Sum also has a strong few-shot video summarization performance.

\paragraph{The Impact of CLIP Prior Mechanism.}
The proposed CLIP Prior mechanism is the key to reduce the domain gap and improve the generalization capacity. To verify the effectiveness of CLIP Prior mechanism, we compare the F1-Scores of removing our CLIP Prior mechanism during pre-training or fine-tuning (we employ weakly-supervised fine-tuning in this experiment) in \cref{tab:ablation_clip_prior}. Obviously, introducing our CLIP Prior mechanism during pre-training significantly improves performance. And CLIP-Prior mechanism is also effective when introduced to fine-tuning. These results demonstrate that the proposed CLIP Prior mechanism is very effective in improving the generalization capacity of the pre-trained Cap2Sum model.

\begin{table}[h]
\centering
\begin{tabular}{@{}l|c|c|c@{}}
\toprule
Method                                        & Publication  & SumMe & TVSum \\ \midrule
SUM-FCN$_{unsup}$ \cite{rochan2018video_sup} & ECCV 2018    & 39.5  & -     \\
UnpairedVSN \cite{rochan2019video}            & CVPR 2019    & 41.6  & 55.7  \\
Cycle-SUM \cite{Cycle-SUM}               & CVPR 2019   & 41.9  & 57.6  \\
ACGAN \cite{he2019unsupervised}               & ACMMM 2019   & 44.5  & 57.8  \\
SumGraph \cite{park2020sumgraph_sup}         & ECCV 2020    & 47.0  & 57.6  \\
ERA \cite{wu2021era}         & BMVC 2021    & 48.8  & 58.0  \\
DSR-RL-GRU \cite{DSR-RL-GRU}    & ICME 2021    & 50.3  & 60.2  \\
CLIP-It \cite{CLIP-It}                        & NeurIPS 2021 & 50.0  & 62.8  \\ 
CA-SUM \cite{CA-SUM}                        & ACM ICMR 2022 & 51.1  & 61.4  \\ \midrule
Cap2Sum$_{usup}$                              & Proposed     & \textbf{54.5}  & \textbf{63.2}  \\ \bottomrule
\end{tabular}
\caption{The comparison of Cap2Sum to unsupervised approaches using Transfer setting, evaluated by F1-Score (\%).}
\label{tab:usup_comp}
\end{table}

\begin{figure*}[h]
    \centering
    \includegraphics[width=0.95\textwidth]{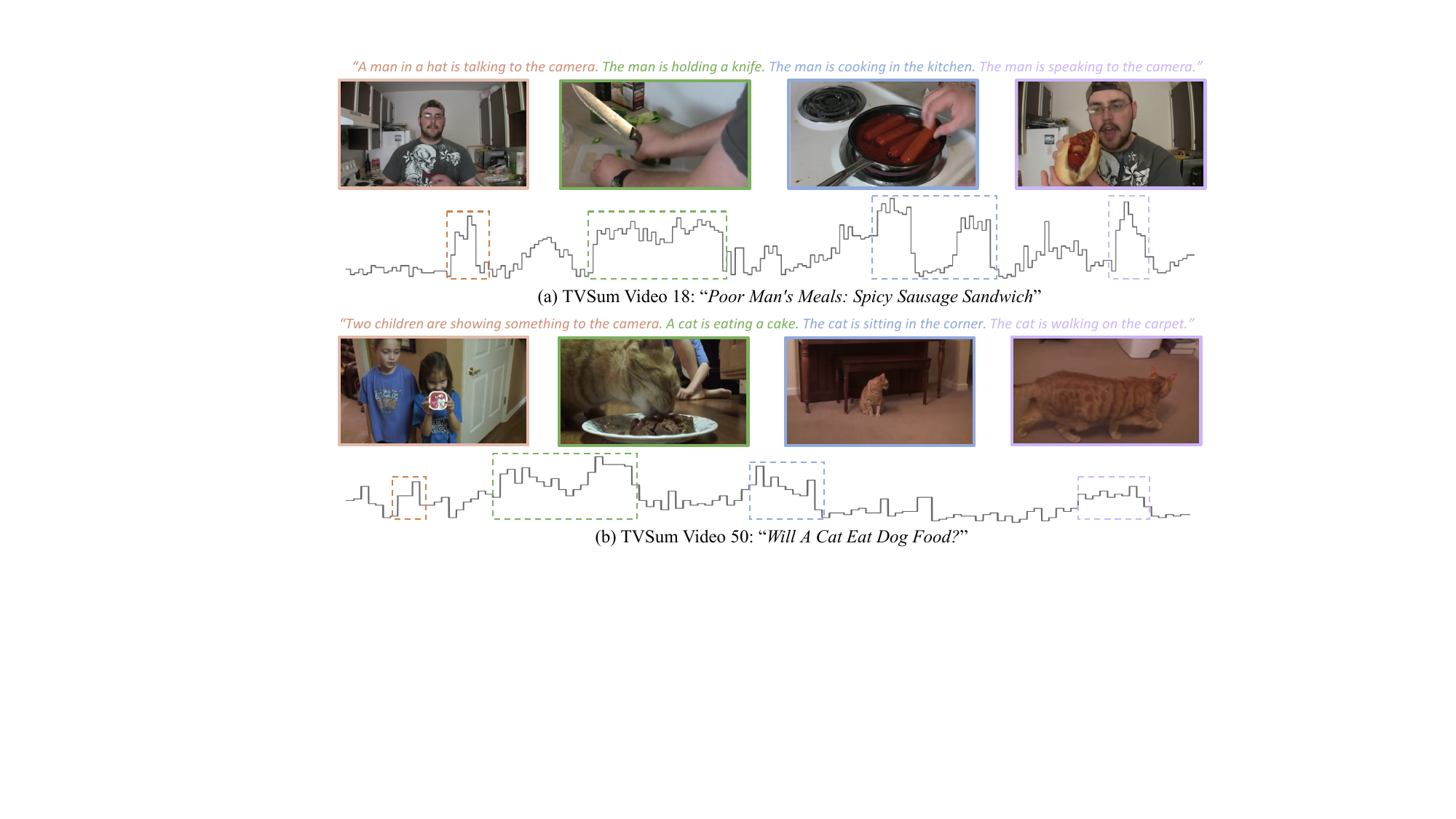}
    \caption{Visualization results with the captions generated by Cap2Sum on TVSum. The gray histogram shows the ground-truth importance scores for each frame and we color-code the corresponding captions, video frames, and importance scores.}
    \label{fig:vis}
\end{figure*}

\subsection{Comparison to Supervised Methods}
In this section, we compare our approach to those state-of-the art supervised video summarization methods. The experiments are conducted using Augmented and Transfer setting, we report the F1-Scores on TVSum and SumMe datasets in \cref{tab:sup_comp}. As the results show, Cap2Sum fine-tuned by ground-truth summarization (\textit{i.e.}, Cap2Sum$_{sup}$) outperforms the state-of-the-art baselines on all settings and on both datasets, thus demonstrating that pre-training using text captions on large-scale datasets is effective in improving the performance. Particularly, compared to CLIP-It \cite{CLIP-It}, which is also a caption-based method, our method achieves significant performance improvement, proving that our approach exploits the video captions in a more efficient way. We also compare the performance of Cap2Sum fine-tuned by video captions (\textit{i.e.}, Cap2Sum$_{wsup}$) and the results show that Cap2Sum also achieves very competitive performance using this setting. Therefore, the effectiveness of learning to summarize videos by generating captions is proven.

\subsection{Comparison to Unsupervised Methods}
Since Cap2Sum is able to be trained without ground-truth summarization, but by other weak labels, similar to the previous unsupervised video summarization methods \cite{khosla2013large,mahasseni2017unsupervised}. It is fair to compare our method (without fine-tuning) with other unsupervised methods. We use the Transfer setting for fair comparison because the videos in the testing dataset are not introduced during the training phase of unsupervised Cap2Sum (Cap2Sum$_{usup}$). We report the F1-Scores in \cref{tab:usup_comp}. In this zero-shot scenario, Cap2Sum outperforms the previous methods on both datasets. We again observe significant performance improvements compared to CLIP-It \cite{CLIP-It} and our Cap2Sum stands state-of-the-art performance among all approaches, which strongly demonstrates the effectiveness of our approach.

\subsection{Visualization}
\cref{fig:vis} shows qualitative results with the captions generated by Cap2Sum on TVSum. We pick 4 video clips with the highest caption confidence and proposal score in each video. We observe that the generated captions have a good consistency with the corresponding video segments and ground-truth importance scores, enabling to train a robust and accurate video summarizer.

\section{Conclusion}
In this paper, we introduce Cap2Sum, a video summarization model learning to summarize videos by generating dense video captions. Our method is the first video summarization approach that can be trained on large-scale datasets. Benefiting from the pre-training on the large-scale datasets, Cap2Sum achieves amazing zero-shot and few-shot video summarization performance on SumMe and TVSum datasets. In addition, we propose a CLIP Prior mechanism and the experiments demonstrate this mechanism is very effective in improving the generalization capacity of Cap2Sum. Generally speaking, the significant performance improvement in zero-shot scenarios makes practical applications of video summarization technology possible. And the proposed weakly-supervised fine-tuning let the users can customize Cap2Sum to generalize the model to their videos. This is also a very useful feature for the video summarization model.

\bibliographystyle{plainnat}
\bibliography{custom}

\begin{thebibliography}{48}
\providecommand{\natexlab}[1]{#1}
\providecommand{\url}[1]{\texttt{#1}}
\expandafter\ifx\csname urlstyle\endcsname\relax
  \providecommand{\doi}[1]{doi: #1}\else
  \providecommand{\doi}{doi: \begingroup \urlstyle{rm}\Url}\fi

\bibitem[Apostolidis et~al.(2020)Apostolidis, Adamantidou, Metsai, Mezaris, and Patras]{apostolidis2020unsupervised_unsup_adv}
Evlampios Apostolidis, Eleni Adamantidou, Alexandros~I Metsai, Vasileios Mezaris, and Ioannis Patras.
\newblock Unsupervised video summarization via attention-driven adversarial learning.
\newblock In \emph{International Conference on multimedia modeling}, pages 492--504. Springer, 2020.

\bibitem[Apostolidis et~al.(2022)Apostolidis, Balaouras, Mezaris, and Patras]{CA-SUM}
Evlampios Apostolidis, Georgios Balaouras, Vasileios Mezaris, and Ioannis Patras.
\newblock Summarizing videos using concentrated attention and considering the uniqueness and diversity of the video frames.
\newblock In \emph{Proceedings of the 2022 International Conference on Multimedia Retrieval}, pages 407--415, 2022.

\bibitem[Cai et~al.(2018)Cai, Zuo, Davis, and Zhang]{DBLP:conf/eccv/CaiZDZ18_wsup}
Sijia Cai, Wangmeng Zuo, Larry~S. Davis, and Lei Zhang.
\newblock Weakly-supervised video summarization using variational encoder-decoder and web prior.
\newblock In Vittorio Ferrari, Martial Hebert, Cristian Sminchisescu, and Yair Weiss, editors, \emph{Computer Vision - {ECCV} 2018 - 15th European Conference, Munich, Germany, September 8-14, 2018, Proceedings, Part {XIV}}, volume 11218 of \emph{Lecture Notes in Computer Science}, pages 193--210. Springer, 2018.

\bibitem[Fajtl et~al.(2018)Fajtl, Sokeh, Argyriou, Monekosso, and Remagnino]{fajtl2018summarizing}
Jiri Fajtl, Hajar~Sadeghi Sokeh, Vasileios Argyriou, Dorothy Monekosso, and Paolo Remagnino.
\newblock Summarizing videos with attention.
\newblock In \emph{Asian Conference on Computer Vision}, pages 39--54. Springer, 2018.

\bibitem[Gygli et~al.(2014)Gygli, Grabner, Riemenschneider, and Gool]{gygli2014summe}
Michael Gygli, Helmut Grabner, Hayko Riemenschneider, and Luc~Van Gool.
\newblock Creating summaries from user videos.
\newblock In \emph{European conference on computer vision}, pages 505--520. Springer, 2014.

\bibitem[Gygli et~al.(2015)Gygli, Grabner, and Van~Gool]{gygli2015video_sup}
Michael Gygli, Helmut Grabner, and Luc Van~Gool.
\newblock Video summarization by learning submodular mixtures of objectives.
\newblock In \emph{Proceedings of the IEEE conference on computer vision and pattern recognition}, pages 3090--3098, 2015.

\bibitem[He et~al.(2023)He, Wang, Qiu, Bui, Shrivastava, and Wang]{A2Summ}
Bo~He, Jun Wang, Jielin Qiu, Trung Bui, Abhinav Shrivastava, and Zhaowen Wang.
\newblock Align and attend: Multimodal summarization with dual contrastive losses.
\newblock \emph{arXiv preprint arXiv:2303.07284}, 2023.

\bibitem[He et~al.(2019)He, Hua, Song, Zhang, Xue, Ma, Robertson, and Guan]{he2019unsupervised}
Xufeng He, Yang Hua, Tao Song, Zongpu Zhang, Zhengui Xue, Ruhui Ma, Neil Robertson, and Haibing Guan.
\newblock Unsupervised video summarization with attentive conditional generative adversarial networks.
\newblock In \emph{Proceedings of the 27th ACM International Conference on multimedia}, pages 2296--2304, 2019.

\bibitem[Ho et~al.(2018)Ho, Chiu, and Wang]{DBLP:conf/eccv/HoCW18_wsup}
Hsuan{-}I Ho, Wei{-}Chen Chiu, and Yu{-}Chiang~Frank Wang.
\newblock Summarizing first-person videos from third persons' points of views.
\newblock In Vittorio Ferrari, Martial Hebert, Cristian Sminchisescu, and Yair Weiss, editors, \emph{Computer Vision - {ECCV} 2018 - 15th European Conference, Munich, Germany, September 8-14, 2018, Proceedings, Part {XV}}, volume 11219 of \emph{Lecture Notes in Computer Science}, pages 72--89. Springer, 2018.

\bibitem[Iashin and Rahtu(2020)]{BMT_Iashin_2020}
Vladimir Iashin and Esa Rahtu.
\newblock A better use of audio-visual cues: Dense video captioning with bi-modal transformer.
\newblock In \emph{British Machine Vision Conference (BMVC)}, 2020.

\bibitem[Jiang and Mu(2022)]{iPTNet}
Hao Jiang and Yadong Mu.
\newblock Joint video summarization and moment localization by cross-task sample transfer.
\newblock In \emph{Proceedings of the IEEE/CVF Conference on Computer Vision and Pattern Recognition}, pages 16388--16398, 2022.

\bibitem[Jung et~al.(2019)Jung, Cho, Kim, Woo, and Kweon]{jung2019discriminative_unsup_adv}
Yunjae Jung, Donghyeon Cho, Dahun Kim, Sanghyun Woo, and In~So Kweon.
\newblock Discriminative feature learning for unsupervised video summarization.
\newblock In \emph{Proceedings of the AAAI Conference on artificial intelligence}, volume~33, pages 8537--8544, 2019.

\bibitem[Jung et~al.(2020)Jung, Cho, Woo, and Kweon]{jung2020global_unsup_adv}
Yunjae Jung, Donghyeon Cho, Sanghyun Woo, and In~So Kweon.
\newblock Global-and-local relative position embedding for unsupervised video summarization.
\newblock In \emph{European Conference on Computer Vision}, pages 167--183. Springer, 2020.

\bibitem[Khosla et~al.(2013)Khosla, Hamid, Lin, and Sundaresan]{khosla2013large}
Aditya Khosla, Raffay Hamid, Chih-Jen Lin, and Neel Sundaresan.
\newblock Large-scale video summarization using web-image priors.
\newblock In \emph{Proceedings of the IEEE conference on computer vision and pattern recognition}, pages 2698--2705, 2013.

\bibitem[Kingma and Ba(2014)]{kingma2014adam}
Diederik~P Kingma and Jimmy Ba.
\newblock Adam: A method for stochastic optimization.
\newblock \emph{arXiv preprint arXiv:1412.6980}, 2014.

\bibitem[Krishna et~al.(2017)Krishna, Hata, Ren, Fei{-}Fei, and Niebles]{anet_caption}
Ranjay Krishna, Kenji Hata, Frederic Ren, Li~Fei{-}Fei, and Juan~Carlos Niebles.
\newblock Dense-captioning events in videos.
\newblock In \emph{{IEEE} International Conference on Computer Vision, {ICCV} 2017, Venice, Italy, October 22-29, 2017}, pages 706--715. {IEEE} Computer Society, 2017.

\bibitem[Krizhevsky et~al.(2009)Krizhevsky, Hinton, et~al.]{cifar100}
Alex Krizhevsky, Geoffrey Hinton, et~al.
\newblock Learning multiple layers of features from tiny images.
\newblock 2009.

\bibitem[Li et~al.(2019)Li, Chen, Zhu, Xie, Huang, Du, and Wang]{Cycle-SUM}
Yanwei Li, Xinze Chen, Zheng Zhu, Lingxi Xie, Guan Huang, Dalong Du, and Xingang Wang.
\newblock Attention-guided unified network for panoptic segmentation.
\newblock In \emph{Proceedings of the IEEE/CVF Conference on Computer Vision and Pattern Recognition}, pages 7026--7035, 2019.

\bibitem[Lin et~al.(2017)Lin, Goyal, Girshick, He, and Doll{\'a}r]{lin2017focal}
Tsung-Yi Lin, Priya Goyal, Ross Girshick, Kaiming He, and Piotr Doll{\'a}r.
\newblock Focal loss for dense object detection.
\newblock In \emph{Proceedings of the IEEE international conference on computer vision}, pages 2980--2988, 2017.

\bibitem[Mahasseni et~al.(2017)Mahasseni, Lam, and Todorovic]{mahasseni2017unsupervised}
Behrooz Mahasseni, Michael Lam, and Sinisa Todorovic.
\newblock Unsupervised video summarization with adversarial lstm networks.
\newblock In \emph{Proceedings of the IEEE conference on Computer Vision and Pattern Recognition}, pages 202--211, 2017.

\bibitem[Narasimhan et~al.(2021)Narasimhan, Rohrbach, and Darrell]{CLIP-It}
Medhini Narasimhan, Anna Rohrbach, and Trevor Darrell.
\newblock Clip-it! language-guided video summarization.
\newblock In Marc'Aurelio Ranzato, Alina Beygelzimer, Yann~N. Dauphin, Percy Liang, and Jennifer~Wortman Vaughan, editors, \emph{Advances in Neural Information Processing Systems 34: Annual Conference on Neural Information Processing Systems 2021, NeurIPS 2021, December 6-14, 2021, virtual}, pages 13988--14000, 2021.

\bibitem[Panda et~al.(2017)Panda, Das, Wu, Ernst, and Roy{-}Chowdhury]{DBLP:conf/iccv/PandaDWER17_wsup}
Rameswar Panda, Abir Das, Ziyan Wu, Jan Ernst, and Amit~K. Roy{-}Chowdhury.
\newblock Weakly supervised summarization of web videos.
\newblock In \emph{{IEEE} International Conference on Computer Vision, {ICCV} 2017, Venice, Italy, October 22-29, 2017}, pages 3677--3686. {IEEE} Computer Society, 2017.

\bibitem[Park et~al.(2020)Park, Lee, Kim, and Sohn]{park2020sumgraph_sup}
Jungin Park, Jiyoung Lee, Ig-Jae Kim, and Kwanghoon Sohn.
\newblock Sumgraph: Video summarization via recursive graph modeling.
\newblock In \emph{European Conference on Computer Vision}, pages 647--663. Springer, 2020.

\bibitem[Phaphuangwittayakul et~al.(2021)Phaphuangwittayakul, Guo, Ying, Xu, and Zheng]{DSR-RL-GRU}
Aniwat Phaphuangwittayakul, Yi~Guo, Fangli Ying, Wentian Xu, and Zheng Zheng.
\newblock Self-attention recurrent summarization network with reinforcement learning for video summarization task.
\newblock In \emph{2021 IEEE International Conference on Multimedia and Expo (ICME)}, pages 1--6. IEEE, 2021.

\bibitem[Radford et~al.(2021)Radford, Kim, Hallacy, Ramesh, Goh, Agarwal, Sastry, Askell, Mishkin, Clark, Krueger, and Sutskever]{CLIP}
Alec Radford, Jong~Wook Kim, Chris Hallacy, Aditya Ramesh, Gabriel Goh, Sandhini Agarwal, Girish Sastry, Amanda Askell, Pamela Mishkin, Jack Clark, Gretchen Krueger, and Ilya Sutskever.
\newblock Learning transferable visual models from natural language supervision.
\newblock In Marina Meila and Tong Zhang, editors, \emph{Proceedings of the 38th International Conference on Machine Learning, {ICML} 2021, 18-24 July 2021, Virtual Event}, volume 139 of \emph{Proceedings of Machine Learning Research}, pages 8748--8763. {PMLR}, 2021.

\bibitem[Rezatofighi et~al.(2019)Rezatofighi, Tsoi, Gwak, Sadeghian, Reid, and Savarese]{giou}
Hamid Rezatofighi, Nathan Tsoi, JunYoung Gwak, Amir Sadeghian, Ian Reid, and Silvio Savarese.
\newblock Generalized intersection over union: A metric and a loss for bounding box regression.
\newblock In \emph{Proceedings of the IEEE/CVF conference on computer vision and pattern recognition}, pages 658--666, 2019.

\bibitem[Rochan and Wang(2019)]{rochan2019video}
Mrigank Rochan and Yang Wang.
\newblock Video summarization by learning from unpaired data.
\newblock In \emph{Proceedings of the IEEE/CVF Conference on Computer Vision and Pattern Recognition}, pages 7902--7911, 2019.

\bibitem[Rochan et~al.(2018)Rochan, Ye, and Wang]{rochan2018video_sup}
Mrigank Rochan, Linwei Ye, and Yang Wang.
\newblock Video summarization using fully convolutional sequence networks.
\newblock In \emph{Proceedings of the European conference on computer vision (ECCV)}, pages 347--363, 2018.

\bibitem[Song et~al.(2015)Song, Vallmitjana, Stent, and Jaimes]{song2015tvsum}
Yale Song, Jordi Vallmitjana, Amanda Stent, and Alejandro Jaimes.
\newblock Tvsum: Summarizing web videos using titles.
\newblock In \emph{Proceedings of the IEEE conference on computer vision and pattern recognition}, pages 5179--5187, 2015.

\bibitem[Song et~al.(2023{\natexlab{a}})Song, Gong, Hu, and Zhao]{DPL}
Zifan Song, Xiao Gong, Guosheng Hu, and Cairong Zhao.
\newblock Deep perturbation learning: enhancing the network performance via image perturbations.
\newblock In \emph{International Conference on Machine Learning}, pages 32273--32287. PMLR, 2023{\natexlab{a}}.

\bibitem[Song et~al.(2023{\natexlab{b}})Song, Zhao, Hu, and Miao]{song2023SPL}
Zifan Song, Cairong Zhao, Guosheng Hu, and Duoqian Miao.
\newblock Learning scene-pedestrian graph for end-to-end person search.
\newblock \emph{IEEE Transactions on Industrial Informatics}, 20\penalty0 (2):\penalty0 2979--2990, 2023{\natexlab{b}}.

\bibitem[Song et~al.(2024{\natexlab{a}})Song, Hu, and Zhao]{song2024diverse}
Zifan Song, Guosheng Hu, and Cairong Zhao.
\newblock Diverse person: Customize your own dataset for text-based person search.
\newblock In \emph{Proceedings of the AAAI Conference on Artificial Intelligence}, volume~38, pages 4943--4951, 2024{\natexlab{a}}.

\bibitem[Song et~al.(2024{\natexlab{b}})Song, Wang, Zhang, Liu, Lyu, Song, Guo, Yan, Lin, Chen, et~al.]{song2024alchemistcoder}
Zifan Song, Yudong Wang, Wenwei Zhang, Kuikun Liu, Chengqi Lyu, Demin Song, Qipeng Guo, Hang Yan, Dahua Lin, Kai Chen, et~al.
\newblock Alchemistcoder: Harmonizing and eliciting code capability by hindsight tuning on multi-source data.
\newblock \emph{Advances in Neural Information Processing Systems}, 37:\penalty0 2185--2214, 2024{\natexlab{b}}.

\bibitem[Song et~al.(2025)Song, Gong, Hu, Dou, Zhao, and Zhao]{song2025dpl++}
Zifan Song, Xiao Gong, Guosheng Hu, Shuguang Dou, Qingsong Zhao, and Cairong Zhao.
\newblock Dpl++: Advancing the network performance via image and label perturbations.
\newblock \emph{IEEE Transactions on Pattern Analysis and Machine Intelligence}, 2025.

\bibitem[Wang et~al.(2018)Wang, Jiang, Ma, Liu, and Xu]{activitynet_caption}
Jingwen Wang, Wenhao Jiang, Lin Ma, Wei Liu, and Yong Xu.
\newblock Bidirectional attentive fusion with context gating for dense video captioning.
\newblock In \emph{2018 {IEEE} Conference on Computer Vision and Pattern Recognition, {CVPR} 2018, Salt Lake City, UT, USA, June 18-22, 2018}, pages 7190--7198. Computer Vision Foundation / {IEEE} Computer Society, 2018.

\bibitem[Wang et~al.(2021)Wang, Zhang, Lu, Zheng, Cheng, and Luo]{PDVC}
Teng Wang, Ruimao Zhang, Zhichao Lu, Feng Zheng, Ran Cheng, and Ping Luo.
\newblock End-to-end dense video captioning with parallel decoding.
\newblock In \emph{2021 {IEEE/CVF} International Conference on Computer Vision, {ICCV} 2021, Montreal, QC, Canada, October 10-17, 2021}, pages 6827--6837. {IEEE}, 2021.

\bibitem[Wang et~al.(2024)Wang, Xu, He, Song, Wang, Qiao, Zhao, et~al.]{wang2024does}
Yi~Wang, Jilan Xu, Yinan He, Zifan Song, Limin Wang, Yu~Qiao, Cairong Zhao, et~al.
\newblock Does video-text pretraining help open-vocabulary online action detection?
\newblock \emph{Advances in Neural Information Processing Systems}, 37:\penalty0 47908--47930, 2024.

\bibitem[Wu et~al.(2021)Wu, Lin, and Silva]{wu2021era}
Guande Wu, Jianzhe Lin, and Cl{\'a}udio~T Silva.
\newblock Era: Entity relationship aware video summarization with wasserstein gan.
\newblock \emph{arXiv preprint arXiv:2109.02625}, 2021.

\bibitem[Xu et~al.(2016)Xu, Mei, Yao, and Rui]{msrvtt}
Jun Xu, Tao Mei, Ting Yao, and Yong Rui.
\newblock {MSR-VTT:} {A} large video description dataset for bridging video and language.
\newblock In \emph{2016 {IEEE} Conference on Computer Vision and Pattern Recognition, {CVPR} 2016, Las Vegas, NV, USA, June 27-30, 2016}, pages 5288--5296. {IEEE} Computer Society, 2016.

\bibitem[Yuan et~al.(2019)Yuan, Tay, Li, Zhou, and Feng]{yuan2019cycle_unsup}
Li~Yuan, Francis~EH Tay, Ping Li, Li~Zhou, and Jiashi Feng.
\newblock Cycle-sum: Cycle-consistent adversarial lstm networks for unsupervised video summarization.
\newblock In \emph{Proceedings of the AAAI Conference on Artificial Intelligence}, volume~33, pages 9143--9150, 2019.

\bibitem[Zhang et~al.(2016{\natexlab{a}})Zhang, Chao, Sha, and Grauman]{zhang2016summary_sup}
Ke~Zhang, Wei-Lun Chao, Fei Sha, and Kristen Grauman.
\newblock Summary transfer: Exemplar-based subset selection for video summarization.
\newblock In \emph{Proceedings of the IEEE conference on computer vision and pattern recognition}, pages 1059--1067, 2016{\natexlab{a}}.

\bibitem[Zhang et~al.(2016{\natexlab{b}})Zhang, Chao, Sha, and Grauman]{zhang2016video_sup}
Ke~Zhang, Wei-Lun Chao, Fei Sha, and Kristen Grauman.
\newblock Video summarization with long short-term memory.
\newblock In \emph{European conference on computer vision}, pages 766--782. Springer, 2016{\natexlab{b}}.

\bibitem[Zhang et~al.(2018)Zhang, Grauman, and Sha]{zhang2018retrospective_sup}
Ke~Zhang, Kristen Grauman, and Fei Sha.
\newblock Retrospective encoders for video summarization.
\newblock In \emph{Proceedings of the European conference on computer vision (ECCV)}, pages 383--399, 2018.

\bibitem[Zhao and Xing(2014)]{Quasi}
Bin Zhao and Eric~P. Xing.
\newblock Quasi real-time summarization for consumer videos.
\newblock In \emph{2014 {IEEE} Conference on Computer Vision and Pattern Recognition, {CVPR} 2014, Columbus, OH, USA, June 23-28, 2014}, pages 2513--2520. {IEEE} Computer Society, 2014.

\bibitem[Zhao et~al.(2021)Zhao, Li, Lu, and Li]{RSGN}
Bin Zhao, Haopeng Li, Xiaoqiang Lu, and Xuelong Li.
\newblock Reconstructive sequence-graph network for video summarization.
\newblock \emph{IEEE Transactions on Pattern Analysis and Machine Intelligence}, 44\penalty0 (5):\penalty0 2793--2801, 2021.

\bibitem[Zhou et~al.(2018{\natexlab{a}})Zhou, Qiao, and Xiang]{zhou2018deep_unsup}
Kaiyang Zhou, Yu~Qiao, and Tao Xiang.
\newblock Deep reinforcement learning for unsupervised video summarization with diversity-representativeness reward.
\newblock In \emph{Proceedings of the AAAI Conference on Artificial Intelligence}, volume~32, 2018{\natexlab{a}}.

\bibitem[Zhou et~al.(2018{\natexlab{b}})Zhou, Xu, and Corso]{youcookii}
Luowei Zhou, Chenliang Xu, and Jason~J. Corso.
\newblock Towards automatic learning of procedures from web instructional videos.
\newblock In Sheila~A. McIlraith and Kilian~Q. Weinberger, editors, \emph{Proceedings of the Thirty-Second {AAAI} Conference on Artificial Intelligence, (AAAI-18), the 30th innovative Applications of Artificial Intelligence (IAAI-18), and the 8th {AAAI} Symposium on Educational Advances in Artificial Intelligence (EAAI-18), New Orleans, Louisiana, USA, February 2-7, 2018}, pages 7590--7598. {AAAI} Press, 2018{\natexlab{b}}.

\bibitem[Zhu et~al.(2022)Zhu, Han, Lu, and Zhou]{RR-STG}
Wencheng Zhu, Yucheng Han, Jiwen Lu, and Jie Zhou.
\newblock Relational reasoning over spatial-temporal graphs for video summarization.
\newblock \emph{IEEE Transactions on Image Processing}, 31:\penalty0 3017--3031, 2022.

\end{thebibliography}

\end{document}